\documentclass[twocolumn, prd, showpacs]{revtex4}
\usepackage{amssymb}

\def\nq{\hspace*{-1em}}

\def\nhq{\hspace*{-0.5em}}
\def\nhh{\hspace*{-0.3em}}
\def\cm{\hspace*{1cm}}

\def\Jl#1#2{{\it #1\/} {\bf #2},\ }

\def\ApJ#1 {\Jl{Astroph. J.}{#1}}
\def\CQG#1 {\Jl{Class. Quantum Grav.}{#1}}
\def\DAN#1 {\Jl{Dokl. AN SSSR}{#1}}
\def\GC#1 {\Jl{Grav. \& Cosmol.}{#1}}
\def\GRG#1 {\Jl{Gen. Rel. Grav.}{#1}}
\def\JETF#1 {\Jl{Zh. Eksp. Teor. Fiz.}{#1}}
\def\JETP#1 {\Jl{Sov. Phys. JETP}{#1}}
\def\JHEP#1 {\Jl{JHEP}{#1}}
\def\JMP#1 {\Jl{J. Math. Phys.}{#1}}
\def\NPB#1 {\Jl{Nucl. Phys.}{B\ #1}}
\def\NP#1 {\Jl{Nucl. Phys.}{#1}}
\def\PLA#1 {\Jl{Phys. Lett.}{#1A}}
\def\PLB#1 {\Jl{Phys. Lett.}{#1B}}
\def\PRD#1 {\Jl{Phys. Rev.}{D\ #1}}
\def\PRL#1 {\Jl{Phys. Rev. Lett.}{#1}}

\def\al{&\nhh}
\def\lal{&&\nq {}}
\def\eq{Eq.\,}
\def\eqs{Eqs.\,}
\def\beq{\begin{equation}}
\def\eeq{\end{equation}}
\def\bear{\begin{eqnarray}}
\def\bearr{\begin{eqnarray} \lal}
\def\ear{\end{eqnarray}}
\def\earn{\nonumber \end{eqnarray}}
\def\nn{\nonumber\\ {}}

\def\nnn{\nonumber\\ \lal }

\def\eql{\al =\al}

\def\dst{\displaystyle}

\def\fracd#1#2{{\dst\frac{#1}{#2}}}

\def\Half{{\fracd{1}{2}}}

\def\e{{\,\rm e}}

\def\diag{\mathop{\rm diag}\nolimits}

\def\const{{\rm const}}

\def\then{\ \Rightarrow\ }

\def\mn{_{\mu\nu}}
\def\mN{_{\mu}^{\nu}}

\def\vac{{}_{\rm (vac)}}
\def\tot{{}_{\rm (tot)}}
\def\matt{{}_{\rm (matt)}}

\def\N{{\mathbb N}}

\def\sph{spherically symmetric}
\def\ssph{static, spherically symmetric}

\begin{document}

\title{Matter sources for a Null Big Bang}

\author{K.A. Bronnikov}
\affiliation{Center for Gravitation and Fundamental Metrology, VNIIMS,
	46 Ozyornaya St., Moscow 119361, Russia;\\
Institute of Gravitation and Cosmology, Peoples' Friendship University
	of Russia, 6 Miklukho-Maklaya St., Moscow 117198, Russia}
\email{kb20@yandex.ru}

\author{O.B. Zaslavskii}
\affiliation{Astronomical Institute of Kharkov,
	V.N. Karazin National University,
	Ukraine, Svoboda Square 4, Kharkov 61077, Ukraine}
\email{ozaslav@kharkov.ua}

\begin{abstract}
  We consider the properties of stress-energy tensors compatible with a
  Null Big Bang, i.e., cosmological evolution starting from a Killing
  horizon rather than a singularity. For Kantowski-Sachs cosmologies, it is
  shown that if matter satisfies the Null Energy Condition (NEC), then (i)
  regular cosmological evolution can only start from a Killing horizon, (ii)
  matter is absent at the horizon, and (iii) matter can only appear in the
  cosmological region due to interaction with vacuum. The latter is
  understood phenomenologically as a fluid whose stress tensor is
  insensitive to boosts in a particular direction. We also argue that matter
  is absent in a static region beyond the horizon. All this generalizes the
  observations recently obtained for a mixture of dust and a vacuum fluid.
  If, however, we admit the existence of phantom matter, its certain special
  kinds (with the parameter $w \leq -3$) are consistent with a Null Big Bang
  without interaction with vacuum (or without vacuum fluid at all). Then
  in the static region there is matter with $w\geq -1/3$. Alternatively,
  the evolution can begin from a horizon in an infinitely remote past,
  leading to a scenario combining the features of a Null Big Bang and an
  emergent universe.

\end{abstract}

\maketitle

\section{Introduction}

  The remarkable discovery that our Universe is accelarating \cite{accel} and
  its explanation, in the framework of general relativity, in terms of the
  so-called dark energy have posed a number of questions. One of them
  concerns the possible interaction between dark energy and usual matter
  that satisfies the standard (weak, strong, null) energy conditions. The
  second one is a relationship between the algebraic structure of the
  stress-energy tensor of dark matter (which is believed to represent
  above 70 per cent of the modern energy density) and possible types of
  evolution. The third one is how these types of evolution depend on the
  possible interaction between the ingredients, and vice versa, how they
  select or restrict the possible kinds of sources.

  The advent of this new and important source of gravity can also shed new
  light on such a long-standing problem of relativistic cosmology as that of
  the initial cosmological singularity. The latter is a state of the
  space-time geometry (and, most frequently, of matter as well) which cannot
  be described in the classical framework. The most common way of its
  understanding is to appeal to quantum gravity, assuming a quantum birth of
  the Universe, but there are a number of interesting attempts to avoid a
  singularity classically or semiclassically. Such attempts can be
  classified as follows: (a) an eternal stationary or quasistationary state
  followed by expansion (the so-called ``emergent universes''), (b) an
  indefinitely long contraction phase followed by a bounce or a number of
  bounces (e.g., nonsimultaneous bounces in different directions in an
  anisotropic Universe), (c) periodic or quasi-periodic evolution, and (d)
  cosmological expansion starting from a Killing horizon, with a static or
  stationary state in the absolute past.

  Let us discuss the fourth opportunity, called a Null Big Bang (or simply a
  Null Bang (see \cite{bd03,bd07} and references therein), using a certain
  phenomenological description of dark energy. The properties of this
  phenomenon were recently discussed in Ref.\,\cite{bd07} among other
  features of the class of regular homogeneous T-models with dustlike matter
  and a vacuum dark fluid. The latter is a variable generalization of the
  cosmological constant, able to account for the present acceleration of the
  Universe, see \eq (\ref{Tvac}) below, and is distinguished by the
  property of invariance under boosts in a particular direction, related to
  symmetry of the model under study \cite{dym92}. It was shown, in
  particular \cite{bd07}, in the framework of Kantowski-Sachs (KS)
  spherically symmetric cosmologies, that, for a mixture of dust and the
  vacuum dark fluid,
\begin{description}
\item[(i)]
   regular cosmological evolution can only begin with a Killing horizon,
\item[(ii)]
   dust is absent at the horizon itself (and it was therefore concluded that
   it is also absent in the static region beyond the horizon) and
\item[(iii)]
   dust can appear in the cosmological region only due to interaction with
   the vacuum fluid.
\end{description}
  It thus follows that this kind of vacuum, whose origin may be related to
  quantum effects of matter fields \cite{dym92,dym02} can, in principle,
  address (though phenomenologically at this stage of the study) three
  important problems simultaneously: that of the nature of dark energy, that
  of the initial cosmological singularity and that of the origin of usual
  matter. It can be added that KS cosmologies are not excluded by modern
  observations if one assumes that their sufficiently early isotropization
  \cite{craw}, and the latter may follow from the process of matter creation
  from vacuum --- see a discussion and some estimates in \cite{bd07}.

  The aim of the present paper is to show that all these three observations
  can be generalized to {\it any\/} kind of matter satisfying the null
  energy condition (NEC) --- normal matter, for brevity. One of the
  motivations for such a study is that the equation of state of matter
  created from vacuum may strongly differ from that of dust. The choice of
  the KS geometry is natural due to its spherical symmetry and possible
  links to black hole physics. There are examples of such black hole
  configurations with phantom matter, which contain expanding universes
  beyond the horizon (``black universes'' \cite{bu1, bu2}). A Null Big Bang
  can, in principle, also occur with other geometries, e.g., those with
  cylindrical or planar symmetries, which may be a subject of future
  studies.

  It should be stressed that the geometric properties of Killing horizons
  (mostly in black hole physics) have been studied in much detail, see,
  e.g., the reviews \cite{car73, car79} and monographs \cite{hell, wald}.
  Strange as it may seem, some much simpler but physically important issues
  concerning the relationship between the properties of a cosmological
  horizon and matter which can support it, evaded attention. Our paper is
  trying to fill this gap.

  The paper is organized as follows. In Section II we obtain analogues of the
  above items (i)--(iii) for any kind of normal matter. Section III briefly
  discusses the possible matter content of a static region in the past of a
  Null Big Bang (more details may be found in \cite{bz2}). Section IV points
  out at possible horizons in an infinitely remote past, and Section V
  summarizes the results.

\section {Kantowski--Sachs cosmology}

  Consider a KS \sph\ cosmology with the metric
\bearr                                                        \label{ds-KS}
	ds^2  = b^2(t) dt^2  - a^2 (t) dx^2 - r^2 (t) d\Omega^2,
\nnn \cm
	d\Omega^2 = d\theta^2 + \sin^2 \theta \, d\varphi^2,
\ear
  supported by a source with the stress-energy tensor
\beq
      T\mN\tot = T\mN\vac + T\mN\matt,
\eeq
  where
\beq                                                        \label{Tvac}
     T\mN\vac = \diag(\rho_v, \rho_v, -p_{v\perp }, -p_{v\perp })
\eeq
  describes a vacuum fluid (defined by the condition $T_0^0\vac =T_1^1\vac$,
  guaranteeing invariance of $T\mN\vac$ under Lorentz boosts in the
  distinguished direction $x$ \cite{dym92}) and
\beq                                                        \label{Tmatt}
     T\mN\matt = \diag (\rho_m, -p_{mx}, -p_{m\perp}, -p_{m\perp})
\eeq
  is the contribution of matter, to be considered below in the most general
  form compatible with (\ref{ds-KS}), as an anisotropic fluid.

  In what follows, it is convenient to use the ``quasiglobal'' time
  coordinate, such that $b = a^{-1}$. The coordinate defined in this way, as
  well as its counterpart in \ssph\ metrics, has very important properties
  \cite{vac1,cold}: it always takes finite values $t=t_h$ at Killing
  horizons that separate static or cosmological regions of space-time from
  one another; furthermore, near a horizon, the increment $t-t_h$ is a
  multiple (with a nonzero constant factor) of the corresponding increments
  of manifestly well-behaved Kruskal-type null coordinates, used for
  analytic continuation of the metric across the horizon. This condition
  implies the analyticity requirement for both metric functions $a^2(t)$ and
  $r^2(t)$ at $t = t_h$.  Though, for our consideration, it is quite
  sufficient to require that these functions belong to class $C^2$ of
  smoothness.

  With this coordinate gauge, the combination of Einstein's equations
  ${0\choose 0}-{1\choose 1}$ reads
\beq
     \frac{2\ddot{r}}{r} a^2 = - 8\pi (\rho_{m} + p_{mx}).  \label{01}
\eeq

  So, let us assume that there is a horizon at some $t = t_h$, such that,
  as $t \to t_h$, $r$ is finite and
\beq                                                        \label{hor}
	a^2(t) \approx a_0 (t-t_h)^{n}, \cm n \in \N,
\eeq
  where $n$ is the order of the horizon. Then, it immediately follows
  from (\ref{01}) and the horizon regularity requirement (which implies
  analyticity of $r(t)$ and, in particular, finiteness of $\ddot r$) that
\beq                                                        \label{p-hor}
	\rho_{m} + p_{mx} \to 0 \quad {\rm as}\quad  t\to t_h.
\eeq

  Now, we will assume $\rho_m \geq 0$ and consider different kinds of
  matter:  the ``normal'' one that respects the NEC,
\beq                                                        \label{NEC}
	T\mn\matt \xi^{\mu} \xi^{\nu} \geq 0, \cm \xi_{\mu} \xi^{\mu}=0,
\eeq
  and the ``phantom'' one that violates it.
  Taking in (\ref{NEC}) the null vectors $\xi^\mu = (a,\ a^{-1},\ 0,\ 0)$
  and ${\bar\xi}{}^\mu = (a,\ 0,\ r^{-1},\ 0)$,
  we obtain two necessary conditions for the validity of the NEC:
\bear
	\rho_m + p_{mx} \geq 0,                           \label{NEC1}
\\
	\rho_m + p_{m\bot} \geq 0.                        \label{NEC2}
\ear

\subsection{Normal matter}

  For normal matter, by definition, \eq (\ref{NEC1}) holds, and consequently,
  according to \eq (\ref{01}), $\ddot r \leq 0$. So we can repeat the
  argument of \cite{bd07}: let the system be expanding ($\dot r > 0$) at some
  $t_{1}$. Then, either $r\to 0$ at some earlier instant $t_{s} < t_{1}$
  (which means a curvature singularity) or the singularity is not reached
  due to a Killing horizon at some instant $t_h > t_s$.

  Thus {\it item {\rm (i) (see the Introduction)} is valid not only for dust
  but for any normal matter.}

  Let us assume that near the horizon the pressure of our matter behaves as
\beq
	p_{mx} \approx w \rho_m, \cm w > -1.                 \label{w>}
\eeq
  (The case $w=-1$ is excluded since it is precisely a vacuum behaviour.)
  Then it immediately follows from (\ref{p-hor}) that {\it both\/}
  $\rho_m \to 0$ and $p_{mx} \to 0$ at the horizon, and the relation
  (\ref{w>}) refers to the first non-vanishing term of a Taylor expansion
  of the function $p_{mx} (\rho_m)$ near $\rho_m =0$. This proves item (ii),
  namely, the statement that normal matter is absent at the horizon.

  Note that both inferences, items (i) and (ii), have been obtained
  irrespectively of whether the normal matter obeys the conservation law
  $\nabla_\nu T\mN\matt =0$ or interacts with a vacuum fluid. (The latter,
  by definition, does not contribute to the expression $T_0^0 - T_1^1$,
  relevant to the NEC.) They also do not depend on the behaviour of $p_{\bot}$
  and on the validity of the energy conditions other than the NEC.

  Our next step is to show that the condition $\rho_m(t_h) = p_{mx}(t_h) =0$
  cannot take place for non-interacting normal matter, which will prove item
  (iii).

  Assuming the absence of interaction between matter and vacuum, the
  conservation law $\nabla_\nu T\mN =0$ should hold for each of them
  separately. Taking the component with $\mu =0$, we obtain
\beq
	\dot{\rho}_{m} + \frac{\dot a}{a}(\rho_m + p_{mx})
	+\frac{2\dot{r}}{r}(\rho_{m} + p_{m\perp })  = 0    \label{cons-m}
\eeq
  and
\beq                                                        \label{cons-v}
	\dot{\rho}_v + \frac{2\dot{r}}{r}(\rho_v + p_{v\perp }) =0.
\eeq

  As to the transverse pressure, we only assume that (at least in
  the limit $\rho\to 0$)
\beq
	|p_{\bot}|/\rho  <  \infty.                       \label{p_bot}
\eeq
  It is a very weak restriction: indeed, for comparison, the dominant energy
  condition would require $|p_{\bot}|/\rho \leq 1$. Then the term in \eq
  (\ref{cons-m}) with $2{\dot r}/r$, which is finite, can be neglected as
  compared with the term containing $\dot{a}/a \sim n/[2(t-t_h)] \to
  \infty$. Therefore, the leading order of the solution to (\ref{cons-m})
  near the horizon reads
\beq
	\rho_m = \const\cdot a^{-(w+1)},                  \label{mhor}
\eeq
  which diverges as $a \to 0$ if $w >-1$, contrary to item (ii).
  Consequently, non-interacting normal matter cannot exist in a KS cosmology
  with a horizon.

  Thus item (iii) has also been proved: normal matter could only appear
  after a Null Big Bang due to interaction with a sort of vacuum.

\subsection{Phantom matter}

  Consider, for completeness, phantom matter with $w <-1$ in \eq (\ref{w>}).
  Again, as with normal matter, the condition (\ref{p-hor}) implies that
  both $\rho_m$ and $p_{mx}$ vanish at the horizon (unless $w$ is variable
  and tends to $-1$ at the horizon, which is a vacuumlike behaviour).
  According to (\ref{mhor}), however, regular solutions to the conservation
  equation (\ref{cons-m}), with zero density and pressure at the horizon, do
  exist.

  The NEC and other energy conditions are violated now. Moreover, \eq (\ref
  {01}) now leads to $\ddot{r}>0$, so that expansion in $r$ can begin from a
  nonsingular state with $r\neq 0$, or $r$ can have a minimum, and the
  presence of a Killing horizon is not necessary for obtaining a nonsingular
  cosmology.

  If there is a Killing horizon, further information on the system behaviour
  near the horizon can be obtained if, in addition to the conservation law,
  we take into account the Einstein equations, of which two independent
  components may be chosen as (\ref{01}) and the ${0\choose 0}$ equation
  that reads
\beq
	\frac{1}{r^{2}}(1+\dot{r}^2 a^2 + 2a\dot{a}r\dot{r})
				= 8\pi(\rho_m+\rho_v).	     \label{00}
\eeq

  Near the horizon, assuming sufficient smoothness of the corresponding
  functions of $t$, we can write the Taylor expansions
\bear
   	a^2 \eql  a_n^2 (\Delta t)^n [1+o(1)],
\nn
   	r(t)\eql r_h+\Delta t\dot{r}_h +
   			\Half {\ddot r}_h (\Delta t)^2 +o(\Delta t)^2,
\nn
  	\rho_m \eql \rho_k (\Delta t)^k [1+o(1)], 	     \label{Tay}
\ear
  where $\Delta t=t-t_h$ and the constants $a_n,\ r_h,\ \rho_k$ are
  positive. Comparing \eqs (\ref{Tay}) and (\ref{mhor}), we see that
\beq
	k=-(w+1)\frac{n}{2} \ \then\ \ w=-1-\frac{2k}{n}.   \label{kn}
\eeq
  It follows from \eq (\ref{01}) that $k\geq n$ where $k>n$ corresponds to
  $\ddot{r}_h = 0$. Thus
\beq
	w \leq - 3,
\eeq
  and, in the generic case $\ddot{r}_h \ne 0$, we have $w = -3$.

  The remaining equation (\ref{00}) gives in the main approximation
  (written for each term separately)
\beq                                                                 
	1 + n a_n^2 (\Delta t)^{n-1} r_h\dot{r}_h =
		8\pi r_h^2 [\rho_v (t_h) + \rho_k (\Delta t)^k].
\eeq
  This leads to two opportunities:

\begin{description}
\item[(a)]
	$\rho_v (t_h) \ne 0$. Then we may have
\bearr
	{\rm either} \ \ n > 1 \ \ {\rm and} \ \
			\rho_v (t_h) = \frac{1}{8\pi r_h^2}
\nnn
        {\rm or}\ \  n=1 \ \ {\rm and} \ \
	\rho_v (t_h) = \frac{1}{8\pi r_h^2 }(1+ a_1^2 r_h\dot{r}_h).
\ear
\item[(b)]
  $\rho_v(t_h) =0$. Then, collecting zero-order terms, we obtain
  $n = 1$ (that is, the horizon is simple, Schwarzschild-like) and
  $\dot{r}_h = -1/(a_1^2 r_h)$ (i.e., a universe appearing in a Null Bang is
  initially contracting in the two spherical directions, $\dot r <0$).
\end{description}

\section{Beyond the horizon: a static region}

  In this paper, we are dealing with essentially nonstatic geometries.
  Meanwhile, if there is a static region separated by a horizon from a KS
  region, one can deduce restrictions on the properties of matter in
  such a region in the same manner. This actually clarifies the conditions
  under which static, spherically symmetric black holes can exist inside
  matter distributions. Such an analysis has been performed in
  Ref.\,\cite{bz2}. It turned out that a regular black hole can be in
  equilibrium with matter having
\[
  	w_r = p_r/\rho = -n/(n+2k) \geq -1/3, \qquad k\geq n,
\]
  where $p_r$ is the radial pressure, $n$ and $k$ are positive integers and,
  as before, $n$ is the order of the horizon. In the generic case $k=n$,
  this gives precisely the value $w=-1/3$ typical of a cloud of disordered
  cosmic strings (see \cite{-1/3} and references therein). Such a cloud,
  however, has an isotropic pressure, which is only a special case in our
  reasoning that uses the radial pressure only.

  We see that, at different sides of a horizon, the kinds of admissible
  matter are different ($w\leq -3$ in the KS region and $w_r \geq - 1/3$ in
  the static one). This looks natural since, as compared with the equation
  of state $p_{mx} = w\rho$, the roles of $p_{mx}$ and $\rho$ now
  interchange, which leads to the substitution $w \longleftrightarrow 1/w$.

  Let us, however, recall that for any normal matter as well as for generic
  phantom matter we have proved the inferences (i)--(iii) formulated in the
  Introduction, and, in the absence of a fluid with $w_r=-1/3$, the static
  region preceding the cosmological evolution should be purely vacuum.
  Its properties should then coincide with those described in Refs.
  \cite{bd03,bd07}. In particular, this static region can be nonsingular; it
  then contains a regular centre with an asymptotically de Sitter geometry
  near it.

\section{A null bang infinitely long ago}

  As was already mentioned in the Introduction, there are several
  cosmologcial scenarios connected with attempts to avoid a singularity
  classically or semiclassically. In particular, there is a variant in which
  the Universe began ins evolution in an infinite past from an almost static
  state with a nonzero scale factor, the so-called ``emergent universes''
  \cite{emerg}, designated as (a) in the list of possibilities (see the
  Introduction).

  We would like to point out here that, in the KS framework, there exists an
  intermediate variant of nonsingular evolution which combines the properties
  of variants (a) and (d) (the latter is the main subject of this paper). We
  mean the situation that one of the scale factors in the metric
  (\ref{ds-KS}), namely, $a(t)$, vanishes as $t\to -\infty$ while the other,
  $r(t)$, remains finite in the same limit, and both timelike and null
  geodesics starting from $t = -\infty$ are complete. This is what can be
  called a ``remote horizon'' in the past, by analogy with remote horizons
  in static space-times mentioned in \cite{cold, bz2}.

  We will illustrate this opportunity with two examples of such a generic
  behaviour as $t\to -\infty$:
\bearr                                                        \label{ex-A}
   {\rm (A)}\quad\ a\approx a_0 \e^{Ht}, \cm r \approx r_0 + B\e^{Ht},
\\ \lal                                                       \label{ex-B}
   {\rm (B)}\quad\ a\approx a_0\biggl(\frac{t_0}{-t}\biggr)^q\nhq , \qquad
     		  r \approx r_0 + r_1\biggl(\frac{t_0}{-t}\biggr)^s\nhq ,
\ear
  where $a_0,\ r_0,\ r_1,\ H,\ t_0,\ s,\ q = \const >0$. Then, carrying out
  an analysis similar to that of Sec. II, we obtain in case (A) $w = -4$ and
  in case (B) $w = -3 - (s+2)/q < -3$. \eq (\ref{00}) then leads in both
  cases to the requirement $\rho_v \to 1/(8\pi r_0^2)$ as $t\to -\infty$.
  Also, in both cases, the conservation equation (\ref{cons-v}) leads to
  $p_{v\bot} \to 0$, i.e., the vacuum stress tensor should asymptotically
  have the structure
\beq
	  T\mN\vac = \diag(\rho_v, \rho_v, 0, 0).            \label{vac-}
\eeq

  Indeed, \eq (\ref{00}) shows that the quantity $Z := 8\pi r^2 \rho_v -1$
  is, in case (A), at most of the order $O(\e^{3Ht})$ and, in case (B),
  $Z = o\big(|t|^{-s-2}\big)$. On the other hand, \eq (\ref{cons-v}) may be
  rewritten in the form $2r{\dot r} p_{v\bot} = -{\dot Z}/8\pi$. Comparing
  the orders of magnitude at both sides of this equation, we obtain
  $p_{v\bot} \to 0$.

  Thus a combination of the Null Big Bang and emergent universe scenarios
  is possible but only under some special conditions: $w \leq -3$ and a
  particular structure of the vacuum stress-energy tensor in the remote past.

\section{Discussion}

  We have considered KS cosmologies with a source representing a mixture of
  a vacuum dark fluid with the stress-energy tensor (\ref{Tvac}) and some
  non-vacuum matter. We have shown the following:

\begin{enumerate}
\item
    In the presence of normal matter, respecting the NEC, regular cosmological
    evolution can begin with a Killing horizon only. Assuming such regularity,
    hence the existence of a horizon, further properties are proven.

\item
    Normal matter is absent at the horizon. Items 1 and 2 are valid
    irrespective of whether or not normal matter obeys the conservation law,
    e.g., whether or not it interacts with the dark fluid.

\item
    Normal non-interacting matter cannot emerge in the cosmological region.
    It can only appear there due to interaction with the dark fluid. This
    restriction is absent for phantom matter that violates the NEC, with
    some particular values of $w = p_{mx}/\rho \leq -3$, the value $-3$
    being generic. Curiously, in the corresponding static region (if any),
    matter should have a non-phantom equation of state with $w \geq -1/3$,
    the value $-1/3$ being generic.

\item
    In the presence of phantom matter with $w \leq -3$, and the
    asymptotic (\ref{vac-}) of the vacuum stress-energy tensor with $\rho_v
    \to \const >0$, the horizon may occur in an infinitely remote past,
    which leads to a scenario resembling that of an emergent universe.

\end{enumerate}

  Concerning configurations with normal matter, we can conclude that the
  static region, preceding a regular KS evolution, should be filled with a
  vacuum fluid only; the latter can provide the existence of a regular
  centre with an asymptotically de Sitter geometry \cite{bd03, bd07}.

  In our reasoning, relying on the asymptotic behaviour of the density
  and pressure near the horizon, we did not assume any particular equation
  of state and even did not restrict the behaviour of the transverse
  pressure except for its regularity requirement. In this sense, our
  conclusions are model-independent. The fact that the very assumption of
  the existence of a cosmological horizon entails a number of rather general
  conclusions resembles, to some extent, the situation in black hole physics
  where the presence of the horizon greatly simplifies the description of
  the system and reduces the number of possibilities.

\vspace*{4mm}

{\bf Acknowledgment.}
We are grateful to Irina Dymnikova for a helpful discussion.
K.B. appreciates partial financial support from Russian Basic Research
Foundation Project 07-02-13614-ofi\_ts.  O. Z. thanks the Inter-University
Centre for Astronomy and Astrophysics (IUCAA), where the revised version of
this paper was finished, for hospitality and a stimulating atmosphere.

\end{document}